\documentclass[12pt,preprint]{aastex}           

   \newcommand{\nd}{\nodata}
   \newcommand{\tnm}{\tablenotemark}
   \newcommand{\gsim}{\rlap{$>$}{\lower 1.0ex\hbox{$\sim$}}}

   \shorttitle{Optical Spectroscopy of K-selected EROs}
   \shortauthors{Yan, Thompson \&\ Soifer}

   \normalsize

\begin{document}

\title{Optical Spectroscopy of K-selected \\
       Extremely Red Galaxies}

\author{Lin Yan}
\affil{SIRTF Science Center, MS~220-6, Caltech, Pasadena, CA~91125}
\email{lyan@ipac.caltech.edu}

\author{David Thompson}
\affil{The Caltech Optical Observatories, Caltech, Pasadena,  CA~91125}
\email{djt@irastro.caltech.edu}

\and 

\author{B. T. Soifer}
\affil{The Caltech Optical Observatories, and}
\affil{SIRTF Science Center, Caltech, Pasadena,  CA~91125}
\email{bts@irastro.caltech.edu}

\begin {abstract}

We have obtained spectroscopic redshifts for 24 sources from a sample
of bright, K-selected EROs using the Keck-I telescope.  These EROs
have high resolution morphologies from HST and were selected with a
median Ks magnitude of 18.7 and $(F814W - K_s) > 4$~mag. (Yan \&\
Thompson 2003).  Among the 24 redshifts, the majority (92\%) are at $
0.9 < z < 1.5$.  We derived the rest-frame J-band luminosity function
at $z_{median} =1.14$. Our result suggests that the
luminosity evolution between bright EROs at $z\sim 1$ and 
the present-day $>$L$^*$ massive galaxies is at most about 0.7~magnitude.
Combining the morphologies and deep spectroscopy
revealed that the 24 EROs have the following properties.  (1) 86\%\ of
the spectra have absorption features from old stars, suggesting that
the dominant stellar populations seen in the rest-frame UV are old
stars. 50\%\ of the sources have pure absorption lines, while the
remaining 50\%\ have emission lines, indicating recent star
formation. We conclude that the color criterion for EROs is very
effective in selecting old stellar populations at $z \sim 1$, and a
large fraction of these systems with prominent old stellar populations
also have recent star formation.  (2) The 12 emission line systems
have the same number of disk and bulge galaxies as in the remaining 12
pure absorption line systems.  We conclude that spectral classes do
not have a simple, direct correspondence with morphological types.
(3) Three EROs could be isolated, pure passively evolving early-type
galaxies at $z\sim 1$.  This implies that only a small fraction
(10\%--15\%) of early-type galaxies are formed in a rapid burst of
star formation at high redshifts and evolved passively since then.
(4) Three EROs have very red continua and pure emission line spectra.
Their redshifts and star formation rates are similar to that of HR~10.
These three sources are potential candidates for dusty starbursts.
(5) We identified three AGNs (13\%), which is consistent with what has
been found by deep Chandra observations.

\end{abstract}

\section{Introduction}

In recent years, significant progress has been made in the study of
Extremely Red Objects (EROs), {\it i.e.} galaxies selected by the
color criteria of either ($I - K_s) > 4$~mag. or ($R - K_s) >5
$~mag. These are the expected colors of a pure passively evolving old
stellar population at roughly a redshift of one.  Several recent
wide field near-IR surveys (McCarthy et al. 2002; Cimatti et al. 2002;
Roche et al. 2002) have provided large datasets for measuring the
surface densities of EROs down to faint K magnitudes and for ERO
clustering analysis (Daddi et al. 2000a; McCarthy et al. 2001). These
surveys found that the ERO angular correlation strength is comparable to that
of low redshift early-type galaxies (Davis \&\ Geller 1976; Willmer,
da Costa \&\ Pellegrini 1998), and about a factor of 10 higher than
that of average field galaxies. These results are partly due
to the narrow redshift range selected by the color criteria.
Additional interpretation is that the EROs at $z \ge 1.0$ are progenitors of
present-day early-type galaxies (McCarthy et al. 2001).

Spectroscopy of EROs provides additional information on their physical
properties. The spectral properties set constraints on their star
formation history and stellar populations.  Similar to the earlier
spectroscopic work on the individual sources HR~10 (Graham \&\ Dey
1996) and CL0939+4713B (Soifer et al. 1999), spectroscopy of a sample
of 30 EROs from the K20 survey (Cimatti et al. 2002) also found that
EROs are a mix of both emission line and absorption line systems, in a
ratio of 1$:$1.  Based on the presence of emission lines, Cimatti et
al. (2002) concluded that 50\%\ of their ERO sample are dusty
star-forming galaxies and the remaining 50\%\ are early-type systems.
This conclusion is based on an assumption that the galaxy spectral
classes correspond to the galaxy morphological classes.

Combining high resolution imaging from HST with ground-based
K-band images, we carried out a morphological study of 115 EROs
selected with ($F814W - K_s) > 4$~mag. (Yan \&\ Thompson, 2003;
hereafter Paper I).  We found that only 30$\pm$5\% of the EROs have
morphologies consistent with a pure bulge or bulge-dominated galaxy
(equivalent to E/S0), while disks comprise 64$\pm$7\% of the sample.
Only 6\%\ of the EROs remained unclassifiable.  Mergers or strongly
interacting systems, which include sources from both classes, make up
17$\pm$4\% of the full sample.  These results suggest that the nature
of EROs are complex.  High resolution morphologies as well as
spectroscopy are needed to better understand this population at high
redshifts.

The heterogeneous nature of EROs selected by their optical/near-IR
colors has also been highlighted by X-ray and sub-millimeter
observations.  Deep Chandra and XMM observations of EROs with $K_s <
20.1$ reveal that roughly 15--20\%\ of EROs are detected in hard
X-rays, consistent with their being obscured Active Galactic Nuclei
(AGN), while roughly another 15\%\ are detected in the soft X-ray band,
and are therefore either low-luminosity AGNs or starbursts (Alexandra et
al. 2002; Brusa et al. 2002).  The 850$\mu$m observations of
relatively bright EROs ($K_s < 19.5$) produced a fairly low detection
rate, only about 20\% (Thompson 2003, private communication; Andreani
et al. 1999; Mohan et al.  2002).  This suggests that ULIRG type
galaxies are relatively uncommon in bright ERO samples.
However, deep near-IR observations (K$_s > 20.5$) of sub-mm selected
sources found that close to half of the 850$\mu$m detections have very
red colors, with $(J - K_s) > $2.6~mag (Frayer et al. 2003).

We found in Paper I that there can be significant differences in the
ERO populations selected by either $(I - K) > 4$~mag. or $(R - K) >
5$~mag.  Both color selection criteria are sensitive to dust
reddened galaxies and systems dominated by old stellar populations at
a redshift of one. However, using the Bruzual \&\ Charlot 
~\citep{BC96} models, we
showed in Paper I (Figure 8) that the EROs selected with ($I - K) >
4$~mag. can include more systems which have somewhat prolonged star
formation (e.g. $\tau = 1$~Gyr galaxies), whereas the ($R - K) >
5$~mag. selection would be biased against these types because 
star formation could contribute enough blue light to make the galaxy drop
out of $(R - K_s) > 5$~mag. selected sample. On another hand, 
$(R - K_s)$ criteria also selects a higher fraction of 
lower-z sources because the Balmer/4000A breaks are still effectively
contributing to the $(R - K_s)$ colors. We should 
point out that ERO is no longer very useful designation, simply because
galaxies selected are highly dependent on what color criteria and photometric
depths of the data. Different color selections inevitably produce different 
galaxy samples with different properties. It is important to
understand these differences before making any comparisons between
surveys using different color selections. 

To complement the morphological study carried out in Paper I, we have
obtained spectroscopic observations for a subset of the EROs from Paper
I.  The goal was to determine the redshift distribution, luminosity functions, 
spectral properties, and to map out the correspondence between the morphologies
and spectral types.  Throughout the paper, we used the cosmology of 
$\Omega_{m} = 0.3$, $\Omega_\Lambda = 0.7$ and H$_0 = 70$~km/s/Mpc.

\section{Observations and Reductions}
\label{obs}

We obtained spectroscopic observations for 36 sources from the
original sample of 115 EROs, which have a median $K_s$ of 18\fm7 and
$(F814W - K_s) > 4$~mag. Because the positions of the EROs in our
sample are distributed all over the sky (see Paper I), the
spectroscopic target fields were primarily selected by their
accessibility during the observing run, with priority given to fields
containing higher numbers of EROs.  No other selection criteria were
imposed on the spectroscopic targets, and we obtained spectra for 
all of the EROs in each target field.  We used the Low Resolution
Imaging Spectrograph (LRIS, Oke et al. 1995) on the Keck\,{\sc i}
telescope on the nights of UT 2002 September 1-4.  The weather
conditions were not photometric and cirrus were present during all
four nights.  The seeing was around $1\arcsec$.  The D680 dichroic and
the 400~l/mm grating blazed at 8500\AA\ in the LRIS red side were used
to optimize wavelength coverage in the critical 8000\,\AA\ --
1\,$\mu$m region, where the strong spectral features, primarily
[\ion{O}{2}] emission and \ion{Ca}{2} H$+$K absorption, are expected
for $z > 1$ galaxies.  In the LRIS blue side, we used the 300~l/mm
grating blazed at 5000~\AA.  Multi-slit masks were set up for all
WFPC2 fields containing two or more EROs, using $1.2\arcsec$ wide
slits.  In order to facilitate sky subtraction, we dithered the
targets along the slits between successive exposures, with offsets of
$2\arcsec$--$4\arcsec$. The integration time for each exposure was
1200 seconds. The total integration time for each multi-slit mask was 
1--1.5~hrs.

The data were reduced using a suite of IRAF scripts written by Daniel
Stern (private communication 2002).  Each slitlet was treated as a
separate, long-slit spectrum. Bias was subtracted by fitting low order
polynomials to the overscan regions.  The blue and red CCDs were read
out using four and two amplifiers, respectively.  The fitting was done
for each overscan region for each amplifier separately.  Sky lines
were subtracted by fitting high order polynomials along the spatial
direction in each two-dimensional (2D) spectrum.  The order of this
polynomial fitting was scaled to the length of the slit, with a
minimum of three for the shortest slits.  A fringe map was made by
median combining the dithered 2D images, excluding the one from which
the fringe map was subtracted, then the fringe map was subtracted
from each of the 2D spectral images.

Cosmic ray removal was done on the individual images using the IRAF
script SZAP from the DIMSUM package{\footnote{Deep Infrared Mosaicking
Software, a package written by Eisenhardt, Dickinson, Stanford and
Ward, available at http://iraf.noao.edu/contrib/dimsumV2}.  First, the
sky lines were modeled with polynomial fitting and taken out from each
image.  Then object spectra were removed by subtracting the
median-combined and smoothed (with a 5x5 boxcar) images.  Cosmic rays
were identified by comparing the flux in the residual images to the
expected noise calculated from the 2D model.  Pixels affected by
cosmic rays were replaced by the values from the model.  The 2D
spectra were not rectified before extracting the 1D spectra.

After the sky line subtraction and cosmic ray removal, all 2D spectra
of a given ERO were shifted and average-combined to create the final
stacked 2D spectrum.  The IRAF task APALL was then used to extract the
1D spectra.  The wavelength solutions were obtained from 3rd order
polynomial fits to the sky lines in the object spectra.  In the LRIS
red side spectra, there were many sky lines for wavelength
calibration, and the wavelength solutions have residual uncertainties
less than 0.3\AA.  In the blue side, there are a fewer lines, with the
residual uncertainties typically less than 1.0\AA.  Flux calibration
was done using long slit spectra of several standard stars.  The
spectra of these bright stars were also used to remove the atmospheric
absorption bands in the red LRIS spectra around 6867\AA\ --- 6944\AA\
(B-band), 7168\AA\ --- 7394\AA\ (b-band), and 7594\AA\ --- 7684\AA\
(A-band).  We note that the D680 dichroic produces a sharp drop of
sensitivity around 6800\AA\ in the spectra.

\section{Results and Discussion}

\subsection{The Redshift Sample}
\label{redshift}

Of the 36 targets observed, redshifts were obtained for 22
extragalactic EROs. We also identified one M star.  Two additional
EROs have redshifts from the Hawaii redshift survey (Cowie et
al. 1996), which makes a total of 24 EROs with redshifts from the
Paper I sample. The source identifications, K-band magnitudes (within
an aperture diameter of 2.5 times the seeing disk), ($F814W - K_s$)
colors, redshifts, quality ranks, emission line fluxes, morphologies,
and spectral features used for the redshift measurements for the 24
EROs are listed in Table~\ref{tab_v2}. The redshift quality rank
describes how secure a redshift measurement is, where ``a'' is for
redshifts measured from multiple, well detected spectral features,
including both absorption and emission lines; ``b'' is for redshifts
measured from multiple, but weak spectral features; and ``c'' indicates redshifts
derived from only a single spectral feature. Figure~\ref{spec} shows 
six representative spectra, 2 with emission lines and 4 with absorption lines, 
from our whole sample. The spectra in this figure are smoothed
with a boxcar of 11~pixels (20~\AA for the LRIS red spectra and
16~\AA for the LRIS blue spectra). Among the six spectra shown 
in this figure, the three redshifts
have quality a and the remaining three have quality b.

Figure~\ref{zhist} shows the redshift distribution.  The median
redshift of the sample is 1.14.  The distribution is strongly peaked
around this redshift, and weakly extends to both lower redshift ($z
\le 0.8$) and higher redshift ($z \sim 1.5$).  The fraction of $z \le
0.9$ interlopers is 8\%\ (2/24) for our redshift sample. This estimate
of low redshift EROs could have a large error because of the small
size of our sample.  The drop-off at the high redshift end of the
distribution is due to fainter apparent magnitudes, redder colors
because of higher K-corrections, and the strong spectral features
moving out of our wavelength range.  Figure~\ref{cmhist} presents the
distributions of the ($F814W - K_s$) color and the $F814W$ magnitude
for both the sources with and without redshifts.  The 13 sources
without redshifts are 0.5~magnitude fainter in $F814W$ and
0.3~magnitude redder in the ($F814W - K_s$) color than the sources
with redshifts.  However, Figure~\ref{zhist} and Figure~\ref{cmhist}
also show that the sources with and without redshifts have overall
similar properties, and that they appear to be from the same parent 
population.

As described in \S\,\ref{obs}, the 36 spectroscopic targets were
selected by their accessibility during the observing run, with
priority given to fields containing higher numbers of EROs; no other
criteria were imposed.  These 36 sources have roughly the same mix in
morphological classes as in the original sample of 115
EROs. Specifically, the spectroscopic sample has 61\%\ (22/36) of
disks or disk dominated (D+DB) galaxies, and 34\%\ (12/36) of bulges
or bulge dominated (B+BD) systems, in comparison with 64\%\ and 30\%\
respectively in the original sample.  

At the median redshift of our sample ($z \sim 1.1$), the morphologies
are based on rest-frame B band light. The recent ERO morphology study
(Moustakas et al. 2003) based on the HST/ACS F850LP (z band) images
found a slightly higher fraction of bulges or bulge dominated galaxies
(40\%) and somewhat lower fraction of disk type of systems
(40\%). This is expected since the F850LP filter samples the light
long-ward of 4000\AA\, and thus is more sensitive to old stars.

Our success rate in measuring redshifts is high in the B+BD
systems (10/12), but much lower in the D+DB galaxies (12/22).  Among
the 13 sources without redshift, the majority (10/13) are disk
dominated galaxies with small bulges (DB); six have edge-on geometry.
A simple reason for the high success rate in measuring redshifts for
B+BD galaxies is because the ($I - K_s) >4$~mag. color limit is
very effective in selecting the bulges or bulge dominated galaxies
with predominantly old stars and little dust at $z \gsim 1$.  Our
spectroscopic data (see \S\,\ref{old_young}) provides further evidence
for this.  The spectroscopic bias against measuring redshifts for DB
type sources could be due to several reasons. Star formation in some
of these systems could be intrinsically weak, thus difficult to
detect, especially for the edge-on disks which have inherently higher
obscuration.  In addition, the oldest stars and younger star forming 
regions form the strong spectral features we use for redshift
identification.  If the integrated light from the DB type EROs are
mostly from intermediate age stellar populations, their spectra would
have weaker spectral lines, thus it is more difficult to measure their
redshifts.

\subsection{The Rest-frame J-band Luminosity Function at $\mathbf{z_{median} = 1.14}$}
\label{lfunction}

Ideally, when we trace the galaxy evolutionary history,
we want to select a specific galaxy population at high redshift, {\it e.g. z=1},
and compare them with the low redshift sources which should have the same
parent population at high redshift. Only such a fair comparison
will give us a unbiased view of the galaxy evolution.
Our redshift sample has a median redshift of 1, and is selected by 
the color of $(I - K_s > 4)$~mag. and $K_s < 18.7$.
These criteria should select most of luminous $>L^*$ galaxies with 
SEDs of old stellar population at $z \sim 1$,
except small number of dusty galaxies at lower redshifts and photometric
errors causing sources scattered in-and-out of our selection.
In principle, the photometric incompleteness could be quantified 
by simulations. However, in this paper, because our redshift sample is 
fairly small, the errors due to small number statistics will dominate
our analyses. Thus, we will not go into the detailed simulation to 
quantify how complete our selection criteria sample the $z \sim 1$ 
luminous galaxies with SEDs of old stellar population.

To understand the relation between the galaxies selected with
$(I - K_s > 4)$~mag. at $z \sim 1$
and present-day massive galaxies, it is instructive to compare
their luminosity functions, which describe the source density
per co-moving volume per absolute magnitude in a certain filter band.
As shown below, we will compare our rest-frame J-band luminosity function
with the present-day J-band luminosity function for galaxies of all types
(Cole et al. 2001), and with the rest-frame J-band luminosity functions 
at z=1.0 measured from the K20 survey (Pozzetti et al. 2003). Here 
we are limited to use the present-day J-band luminosity function 
for all galaxies types because there are no published data for 
early type galaxies. However, because the sources in our redshift sample
are very luminous, our comparison is most limited to bright end 
of the luminosity function (as seen below). Therefore, the mismatch
is probably very small.

At the median redshift of our sample ($z_{median} = 1.14$), 
the observed $K_s$ filter corresponds to the rest-frame $J$ band.
We compute the absolute magnitude at the rest-frame J-band for 
an object with an apparent K$_s$ magnitude at redshift of z, using the
following equation:

\begin{equation}
\label{absmag}
M_{J} = m_{Ks} - 25 - 5.0\log_{10}(D_{L}/Mpc) - k_c - A
\end{equation}

Here A is the galactic extinction, which is
generally very small, less than 0.04~magnitude in K$_s$ band.
The k-correction term, $k_c$, describes the color difference
between the observed and the rest-frame bandpasses (Oke \&\ Sandage 1986;
Kim, Goobar \&\ Perlmutter 1997). 

\begin{eqnarray}
\lefteqn{k_c =} & & \mbox{}-2.5\log_{10}\left({\int Z(\lambda) S_{J}(\lambda) d\lambda \over \int Z(\lambda)S_{K_{s}}(\lambda) d\lambda} \right) \nonumber \\
& & \mbox{}+2.5\log_{10}(1+z) \nonumber \\
& & \mbox{}+2.5\log_{10}\left({\int F(\lambda) S_{J}(\lambda) 
d\lambda \over \int F(\frac{\lambda} {(1+z)}) S_{K_{s}}(\lambda) d\lambda}\right) 
\label{kcorr}
\end{eqnarray}

Here $Z(\lambda)$ is the SED for $\alpha$ Lyrae,  $S(\lambda)$
is the filter transmission curve, $F(\lambda)$ is the assumed
galaxy SED template.
As shown above, $k_c$ includes the zero point difference between
two different filters, as well as
the color difference due to the redshifted SED
(Kim, Goobar \&\ Perlmutter 1997). At zero redshift, $k_c$
becomes the $J - K_s$ color of a galaxy with the assumed SED.
We took the $\alpha$ Lyrae SED from
Bruzual \&\ Charlot (1996) and the standard Mauna Kea
$J$ and $K_s$ filter curves from Simons \&\ Tokunaga (2002).  
The k-correction was computed for an early-type SED, and is -1.58
at $z = 1.0$. The uncertainties in 
k-correction due to different SEDs should be very small 
because at $z \sim 1$ the observed K$_s$ band is roughly equivalent to 
the rest-frame $J$ band.

We used the 1/V$_{max}$ method (Schmidt 1968; Felten 1976) to calculate
the rest-frame J-band luminosity function. We adopted the cosmology
with $\Omega_{M} = 0.3, \Omega_{\Lambda} = 0.7$ and H$_0=70$~km/s/Mpc.
The maximum co-moving volume for a source 
is defined as the volume within which the source could still be 
above the flux limit of the sample. 
In the 1/V$_{max}$ approach, the galaxy luminosity function
$\phi(M)$, the number of galaxies per co-moving volume per absolute
magnitude, is simply:

\begin{equation}
\phi(M) = {1 \over \Delta M} {\sum_{i} {1 \over V_{max}^i}}
\end{equation} 

For each source within our redshift sample, we compute its maximum volume.
Because of field-to-field differences in exposure times, 
atmospheric seeing, or dithering, our survey area is a function of 
the limiting magnitude. To account
for this effect when calculating the volume factor used in the
luminosity function, we integrated the survey area over the ten
spectroscopic target fields as a function of magnitude (see Paper I for
details).
The resulting differential area histogram was then used to calculate
an effective survey volume for each ERO with a spectroscopic redshift.

Figure~\ref{lf} shows the rest-frame J-band luminosity function of 
the K-selected EROs at $z \sim 1$. The solid points are the measurements
using only the sources with redshifts, whereas the circles are
the measurements of the luminosity function including also the sources
without any redshifts, by artificially setting their redshifts 
to the median redshift of 1.14. This is one of the crude ways to 
compute the incompleteness correction. Because our color criteria
select galaxies within a relatively narrow redshift space, 
setting $z = 1.14$ for galaxies without redshift measurement 
is not a far off approximation. The purpose of doing this is to
demonstrate the approximate amplitude of the correction, rather than
producing an accurate measurement.

In Figure~\ref{lf}, we compare our results with 1). the local
J-band luminosity function (Cole et al. 2001) (solid curve in Figure~\ref{lf}); 
2). the rest-frame J-band luminosity function at $0.75 < z < 1.3$ 
using a K-band flux limited spectroscopic sample (long dashed curve 
in Figure~\ref{lf}) (Pozzetti et al. 2003); 3). the rest-frame
J-band luminosity function at $0.75 < z < 1.3$ for
galaxies with pure absorption lines 
in the K20 survey (dotted curve in Figure~\ref{lf}, Pozzetti 2003,
private communication). It is not surprising that 
our data points are somewhat higher than the dotted curve which 
indicates the LF for pure absorption line galaxies from the K20 survey. 
As shown by our results, selecting only pure absorption line galaxies
will underestimate the volume density of old stellar populations 
at redshift of 1.0.
Ideally, we should compare our result with the local J-band LF of
early type galaxies, however, there is no published measurement
in the literature. As discussed in the begining of this section, 
since our sources are mostly luminous ($>L^*$) galaxies, 
the comparison with the local J-band LF for all galaxy types should 
not be too far off. The luminosity 
evolution between the local LF and the LF at z=1.0 from the K20 survey 
is roughly 0.7~magnitude (Pozzetti et al. 2003). Our measurements lie
roughly along the dashed curve. The EROs are a color selected
subset of a general K-band flux limited sample. The high volume density
implied by our data points can be explained by the significant clustering 
in our ERO redshift sample, since
the spectroscopic targets were selected
with higher priority given to fields containing higher numbers of EROs.
We computed the ERO surface density in our redshift sample, which
is a factor of 2--3 higher than that of the original sample.
This is a direct evidence that the EROs in our redshift sample 
are clustered within small volumes.

Although the rest-frame J-band luminosity function at $z =1.14$
we derived from our sample is highly biased, it does set an upper limit
on the amount of luminosity evolution EROs could have if
compared to the present-day $>$L$^*$ luminous galaxies.
Any properly selected ERO sample should be less clustered
over a larger area, and its luminosity function would be lower
compared to what we measured. Thus, we conclude that the luminosity
evolution between bright EROs at $z \sim 1$ and the present-day massive
galaxies is small, about 0.7~magnitude.

We also compute the rest-frame J-band luminosity density 
from our data, which is $\ell = \sum_i L_i/V_{max}^i$, over the range
of M of -24.0 to -26.0. The logrithmic of the rest-frame J-band luminosity 
density for galaxies brighter than -24, $\log_{10}(\ell )$, 
is 19.86~Watt/Hz/Mpc$^3$ and 20.02~Watt/Hz/Mpc$^3$ for
the redshift sample and the sample including sources without redshifts 
respectively. These numbers are slightly higher than the 
luminosity density computed from the rest-frame J-band LF at $z = 1.0$
from the K20 survey, integrated from -24.0 to -26.0 
($\log_{10}(\ell) = 19.8$~Watt/Hz/Mpc$^3$). Again, this could be explained by
somewhat higher clustering in our small sample. 
In order to evaluate the amplitude of the evolution strength, we
also compute the local J-band luminosity density, $\log_{10}(\ell )$ is 
19.18~Watt/Hz/Mpc$^3$, using the measurement
from Cole et al. (2002). We found that luminosity density evolution between
now and redshift of 1 is $\Delta \log_{10}(\ell )/\Delta \log_{10}(1+z) = 2.3$.

\subsection{Correspondence between Morphologies and Spectral Types}
\label{relation}

[\ion{O}{2}]~$\lambda$3727\AA\ emission was detected in 12 of the 24
sources listed in Table 1.  Thus, 50\%\ of our sample with redshifts
have emission lines, and the other 50\%\ are pure absorption line
galaxies.  These percentages are the same as found in the K20 survey
(Cimatti et al. 2002). In Paper I, we classified disks or disk
dominated systems as D+DB, and bulge or bulge dominated as B+BD.  The
two spectral classes, emission line versus absorption line, have a
similar mix of bulge and disk dominated morphological types.
Specifically, among the 12 sources with [\ion{O}{2}]~$\lambda$3727\AA\
emission lines, 7 are D+DB and 5 are B+BD systems, while of the 12
absorption line sources, 7 are classified as D+DB and 5 as B+BD.  The
ERO morphological classes do not have a simple correspondence with the
presence of emission lines or absorption lines.  As stated in
\S\,\ref{redshift}, the redshift sample (24 sources) have slightly
higher fraction of B+BD galaxies than that of D+DB, on a level of
10\%. Considering our small size sample and large uncertainties, we do
not attempt to correct for any morphological biases in the redshift
sample.

This result is visually shown in Figure~\ref{mosp} using the HST/WFPC2
stamp image of each ERO.  The four panels in this figure present the
four categories of EROs --- (a) disks with emission lines, (b) disks
with pure absorption lines, (c) bulges with emission lines, and (d)
bulges with pure absorption lines.  The emission line sources,
including both bulges and disks (groups a and c), are at slightly
higher redshifts and have slightly brighter $K_s$ magnitudes than the
absorption line systems (groups b and d).  The bulges with pure
absorption lines are the reddest subset in the ($F814W - K_s$) color,
but the two reddest systems are disks.

Figure~\ref{mosp}.a shows the disks with emission lines.  These
systems are probably dust reddened, and also have star forming
activity in their disks. One obvious example of dust reddening is
shown in the second stamp image in panel a; the dust lane is clearly
present in this galaxy.  Figure~\ref{mosp}.b shows the EROs which are
morphologically classified as disk dominated but without detectable
emission lines.  These are the quiescent disks.  In Paper I, we
classified four of these as edge-on disks. It is possible that the
inherently high obscuration in the edge-on orientation make the disks
faint, while the central bulges are relatively unobscured and
bright. This would also explain why these sources are at $z \sim 1$
since the ($I - K_s) > 4$~mag. criterion should select old stellar
populations at these redshifts.  Figure~\ref{mosp}.c presents
``rejuvenated old stellar populations''. These are the bulge dominated
systems (B+BD) with emission lines. The HST/WFPC2 images show that
these bulges with emission lines have clear indications of
mergers/interactions, such as tidal tails, asymmetric nuclei or
multiple faint companions.  The star formation triggered by these
interactions produces the detected emission lines. However, the
dominant stellar populations in these bulges are old stars.
Figure~\ref{mosp}.d presents the five absorption line bulges (B+BD).
Among them, three are pure bulges (no evidence of disks) which could
be identified as passively evolving early-type galaxies at $z \sim 1$.
The statistics (3/24) from our survey suggests that the fraction of
this type of system is small, around 10$-$15\%.  This result sets a
strong constraint on the traditional galaxy formation scenario, in
which massive ellipticals formed at high redshifts in a burst of star
formation and passively evolved to the present day (Kauffmann \&\
Charlot, 1998).

Excluding the three AGNs (see \S~\ref{agnclst}) and the two sources
without reported line fluxes from the Hawaii Redshift Survey (Cowie et
al. 1996), the [\ion{O}{2}]~$\lambda$3727\AA\ line fluxes in the
remaining seven sources are in the range of $5\times 10^{-18}$ ---
$4\times 10^{-17}$~ergs/s/cm$^2$. These line fluxes are the raw
measurements, not corrected for light losses due to the finite slit
width or extinction. Using the conversion from Kennicutt (1998), where
the star formation rate (SFR, M$_\odot$ yr$^{-1}$) = 1.4$\times
10^{-41}$L[\ion{O}{2}]~(ergs/s), we calculate that the SFRs in these
seven sources are in the range of $0.4$ -- $6$~M$_\odot$/yr.  The 
star formation rates measured from the rest frame
[\ion{O}{2}]~$\lambda$3727\AA\ line are only lower limits because dust
extinction could be significant.  The highest star formation rate in
our sample is 6~M$_\odot$/yr for ufg00\_121, comparable to HR~10 (Dey
et al. 1999), which has an [\ion{O}{2}] star-formation rate of
4.8~M$_\odot$/yr in our assumed cosmology model.  HR~10 is a massive
dusty starburst based on millimeter and sub-mm observations,
suggesting that ufg00\_121 could be a similar system. Clearly longer
wavelength observations using SIRTF and SCUBA are required to measure
the total star formation rates for the very dusty systems.

The global volume-averaged star formation rate is shown to rise
steeply from the present day to redshifts of 1 --- 2 (Madau et
al. 1996; Lilly et al. 1996; Yan et al. 1999).  The morphological
properties observed in our ERO sample and the high fraction of star
forming systems could be largely due to this evolutionary effect at $z
\sim 1$.  Finally, our observed result could also be due to the color
selection. The ($I - K_s) >4$~mag. color selects more old, red
bulges with small disks or residual star formation (S0, Sa like
systems) than what is selected with an $(R - K_s) > 5$~mag.
criterion. This is mainly because the $R$-band is more sensitive to
star formation.  More details about this color selection effect are
given in Paper I.

\subsection{Old versus Young Stars}
\label{old_young}

The presence of nebular emission lines, such as
[\ion{O}{2}]~$\lambda$3727\AA\, in spectra are a sensitive probe of
recent star formation.  Although half of our redshift sample have
[\ion{O}{2}] emission, the spectra of most sources have not only very
red continua but also absorption line features, such as H+K, G-band,
or a 4000~\AA\ break.  These features are the indicators of old stars.
Specifically, excluding two EROs from the Hawaii Redshift Survey
(Cowie et al. 1996), we have 19 spectra which show absorption lines,
seven of which also have emission lines. Only three EROs have pure
emission line spectra.  This implies that a large fraction (19/22) of
the EROs in our redshift sample, including both bulge and disk
dominated systems, have mostly old stellar populations with an
additional small amount of young stars formed more recently.  A good
example is shown in the bottom panel of Figure~\ref{spec}, u2fq1\_085. 
The morphology of this ERO
shows a prominent bulge with a faint disk; the red portion of its
spectrum clearly shows absorption features from old stars in the bulge
and the blue portion of its spectrum has an
[\ion{O}{2}]~$\lambda$3727\AA\ emission line, indicating star forming
activity in the disk.  Early-type galaxies with recent star formation
are also found in the field (Treu et al. 2002; van Dokkum \&\ Stanford
2002).  Thus, using only the presence of emission lines as the
diagnostic to separate EROs into the two categories of early-type
versus dusty star forming systems can lead to significant
mis-classification.

At the other extreme, three EROs, ufg00\_121, u2fq1\_130 and u2fq1\_115, 
have only [\ion{O}{2}]~$\lambda$3727\AA\ emission, and no detectable 
absorption lines.  The red continua observed in these three 
sources are probably due to dust extinction.  Two of these EROs have 
disk type morphologies, the other is a bulge. They are the most likely
candidates for being very dusty and having strong star formation like
HR~10 (Dey et al. 1999).  [\ion{O}{2}]~$\lambda$3727\AA\ emission was
detected in HR~10, with an uncorrected SFR of 4.8~M$_\odot$/yr. Its
spectrum shows a very red continuum but no significant absorption
lines.  

Figure~\ref{newcz} shows the ($F814W - K_s$) color as a function of
the redshift for the sample listed in Table 1.  In the figure, the
model curves are calculated using the Bruzual \&\ Charlot (1996)
spectral energy distributions (SED), with different assumptions on 
the star formation history.  The solid dots identify the three bulges 
with absorption line spectra.  These systems could be pure passively 
evolving early-type galaxies formed in a rapid star burst at high 
redshift, evolving only passively since then.  Since their colors are 
slightly redder than the $\tau=0.1$~Gyr models, this could suggest 
that there is a small amount of dust in these galaxies.  Our data 
alone can not discriminate the models with two very different
formation redshifts.

The open circles in the figure indicate the remaining 88\%\ (21/24) of
the sample, which have large or small disks, as well as the bulges 
with emission lines. The observed colors of these EROs are consistent 
with the colors from the three models, where dust, multiple epochs of 
star formation, and old stellar populations all could play some role
in determining their colors.  This highlights the complex physical
properties of EROs at $z \sim 1$.

Ultimately, the goal is to obtain more quantitative estimate of what fraction
of the K-band light is due to 
old stars in these optical/near-IR color selected galaxies. In principle, 
this can be done if one has high S/N spectra and good broad band 
photometry. Stellar population synthese models could be used to fit 
both broad band SEDs, as well as emission line and absorption line spectra.
This will decompose the stellar population and quantitatively measure
the fraction of K-band light due to old stars. Although our current
data does not permit us doing this analyses, in the future, such analyses
should be done properly with better datasets.

\bigskip

\subsection{AGNs and Clusters}\label{agnclst}

Three EROs in our redshift sample have AGN signatures.  Both
u2fq1\_128 and u2fq1\_389, show \ion{C}{3}]~$\lambda$1909\AA\ and
\ion{C}{2}]~$\lambda$2326\AA\ semi-forbidden emission lines in their
spectra. Figure~\ref{agn} shows the spectrum of u2fq1\_128, smoothed
with a boxcar of 5~pixels (7.3\AA).  The observed full widths at half
maximum (FWHM) in the emission lines are about 3740~km/s for
u2fq1\_128 and 790~km/s for u2fq1\_389.  The source u2fq1\_128 is
classified as a bulge dominated system, while u2fq1\_389 has a faint
disk (Paper I).  Both are likely dust-shrouded, low-luminosity AGN,
with the broad-lined u2fq1\_128 being a Seyfert I galaxy and the
narrower-lined u2fq1\_389 being a Seyfert II or LINER galaxy.

The third AGN is ufg00\_044.  The top panel in Figure~\ref{agn2} shows
the HST/WFPC2 F814W images of ufg00\_044 and ufg00\_083. These two
EROs are separated by 6.6 arcseconds (56~kpc projected separation),
both have [\ion{O}{2}]~$\lambda$3727\AA\ emission lines and are at $z
\sim 1.33$ (Figure~\ref{mosp}, bottom left panel).  These two EROs 
are clearly situated in
a high density region, with evidence for ongoing
merging/interaction. The slit went through both ERO ufg00\_044 and
ufg00\_083, as well as two fainter galaxies.  In Figure~\ref{agn2},
the bottom right panel shows the 2D spectra in the observed wavelength
region of 8635 --- 8748\AA, and that ufg00\_044 has a broad
[\ion{O}{2}]~$\lambda$3727\AA\ line. The observed velocity width is
609~km/s. In addition, it also has [\ion{Ne}{3}]~$\lambda$3867\AA\
emission (not shown in Figure~\ref{agn2}), confirming the bright line
is [\ion{O}{2}]~$\lambda$3727\AA. Object ufg00\_044 is clearly in an
actively merging environment. The velocity extension above its broad
[\ion{O}{2}] line probably comes from the close companion (C).
Perhaps this close interaction is the trigger of its AGN activity.
The bottom left panel in Figure~\ref{agn2} covers the wavelength from
7614 --- 7727\AA; the emission line source with no detected continuum
between the two EROs is object F. Its spectrum has two emission lines,
H$\beta$ and [\ion{O}{3}]~$\lambda$5007\AA, and is a foreground galaxy
at $z = 0.50$. The [\ion{O}{3}]~$\lambda$4959\AA\ falls within the
atmospheric absorption A band, and thus this line is only weakly
detected.

The one mega-second Chandra observation in the HDF revealed that about
$14^{+11}_{-7}$~\%\ of $I - K > 4$~mag. selected EROs are detected in the
hard X-ray band and are AGNs (Alexander et al 2002). We detect 3 AGNs
out of 24 EROs with redshifts, which is consistent with the X-ray
studies.  Two of these AGNs have ($F814W - K_s$) colors of 4.10 and
4.16, and are both bulge dominated; the third AGN has an ($F814W -
K_s$) color of 5.25 and disk morphology. The red colors from these
three AGNs could be due to a combination of dust extinction and the
presence of old stellar populations.

The sources observed are in ten different WFPC2 fields (Table 1), each
of which is slightly less than $160\arcsec \times 160\arcsec$.  Among
these ten fields, four contain EROs clustered in redshift space,
likely tracing the high concentration regions of large scale
structures.  The two AGNs, u2fq1\_128 and u2fq1\_389 are at similar
redshifts, 1.467 and 1.466.  In the u2845 field, we find three sources
at $z \sim 0.9$. One is an ERO, while the other two are serendipitous
detections at similar redshifts.  In the ufg00 field, a group of three
EROs ufg00\_083, ufg00\_044, and ufg00\_121, are identified at $z \sim
1.3$.  All three have [\ion{O}{2}]~$\lambda$3727\AA\ emission lines,
and their morphologies show clear signatures of recent
interactions. In the uim03 field, sources uim03\_100 and uim03\_089
are at $z \sim 1.1$.  Although these EROs don't necessarily belong to
clusters of galaxies at $z \sim 1$, they are clearly situated in high
density environments.  These sources do not show a preference for any
morphological or spectral types. The magnitudes, colors, morphologies,
and spectral properties presented in this paper and Paper I suggest
that the $K$-selected EROs are probably massive, luminous systems at
$z \sim 1$. These systems tend to be strongly clustered, tracing peaks
of the mass density field, as found by other wide field surveys (Daddi
et al. 2000a; McCarthy et al. 2002; Roche et al. 2002).

\section{Summary}

We obtained deep spectroscopic observations for 36 sources among the
original 115 EROs, which were selected in Paper I with ($F814W - K_s)
> 4^{\rm m}$ and have a mean $K_s \sim $18\fm7.  The major conclusion of
Paper I is that a large fraction of EROs (64\%) are disks or disk
dominated galaxies, while only 30\%\ have bulge or bulge dominated
morphologies.  Of the 36 spectroscopic targets, we were able to
determine redshifts for 22 sources.  We found that half of the
redshift sample have [\ion{O}{2}]~$\lambda$3727\AA\ emission, while
the remaining half are pure absorption line systems.  The two broad
spectral classes (emission versus absorption line) have an equal mix
of bulge and disk dominated galaxies. If we include the sources
without redshifts (the total sample of 36), the fractions of emission
line and the pure absorption line systems are 1/3 respectively. 
It appears that the spectral
class does not have a simple, direct correspondence with the
morphological type.  Detecting [\ion{O}{2}]~$\lambda$3727\AA\ emission
in 50\%\ of the redshift sample implies that recent star forming
activity is fairly common among the EROs.  However, we
found that although 50\%\ of the sample have star formation, there are
significant populations of old stars in the majority of the sample.
This is supported by the evidence that a large fraction of the spectra
(19/22) in our sample not only have very red continua, but also
absorption features which are indicative of old stars.  Only three
EROs have pure emission line spectra, with no significant absorption.  
They are the most likely candidates for dusty starbursts.

The results from this paper and Paper I have significant implications
for the theories of galaxy formation and evolution.  First, the strong
ERO clustering measured by recent wide field near-IR surveys (McCarthy
et al. 2001; Daddi et al. 2000a) are generally interpreted as the
evidence that EROs selected by a single optical/near-IR color are the
progenitors of present-day ellipticals. However, our detailed studies
using HST/WFPC2 images and Keck spectroscopy show that the properties
of these EROs are fairly complex; a significant fraction of EROs have
disks and/or signs of mergers/interactions.  For these EROs to evolve
to present-day ellipticals, major mergers would be required in order
to re-distribute old stars, and consume all the gas in the disks on a
time scale of a few billion years (from $z \sim 1$ to the present-day)
(Kormendy \&\ Sanders 1992; Shioya \&\ Bekki 1998).  This implies that
most of the EROs would still need to go through a starburst
phase. Sensitive far infrared observations from SIRTF will be very
useful to measure the volume densities of dusty starbursts between $z
\sim 1$ and the present-day.  This will allow us to draw more
quantitative connections among EROs, dusty starbursts and the
present-day early-type galaxies.

Second, the fraction of isolated, passively evolving old stellar
populations at $z \sim 1$ is small, roughly 10$-$15\%.  Clearly the
generic scenario, where massive ellipticals are rapidly formed at high
redshifts (Eggen, Lynden-Bell \&\ Sandage 1962) and then evolve
passively, should be limited to only a small fraction of
the entire early-type galaxy population.  Previous measurements of
the number density of pure passively evolving old stellar populations
based on a single color selection or optical spectroscopic classification 
(Daddi, Cimatti \&\ Renzini, A. 2000b) are probably over-estimated.

The rest-frame J-band luminosity function derived from our sample
suggests that the luminosity evolution between
EROs at $z\sim 1$ and the present-day massive galaxies is probably not very
large. More accurate measurements would require 
a larger and more complete
redshift sample with high resolution images from HST. There exists several
deep and large datasets useful for this.  A complete
redshift survey of these datasets would provide important information
for understanding the nature of the EROs.

\section{Acknowledgment}

We thank Daniel Stern for his generous help and providing his software
for the LRIS data reduction.  D. Thompson acknowledges partial support
from the NASA LTSA grant NAG5-10955. LY and DT acknowledges the
funding for this research by the HST grant AR-08756.  LY and BTS are
supported by the SIRTF Science Center at Caltech.  SIRTF is carried
out at the Jet Propulsion Laboratory, operated by California Institute
of Technology, under contract with the National Aeronautics and Space
Administration. The spectroscopic data presented herein were obtained
at the W.M. Keck Observatory, which is operated as a scientific
partnership among the California Institute of Technology, the
University of California, and the National Aeronautics and Space
Administration.  The Observatory was made possible by the generous
financial support of the W.M. Keck Foundation. We also wish to
recognize and acknowledge the very significant cultural role and
reverence that the summit of Mauna Kea has always had within the
indigenous Hawaiian community.  We are most fortunate to have the
opportunity to conduct observations from this mountain.


\clearpage
\begin{figure}[!thp]
\epsscale{1.1} \plotone{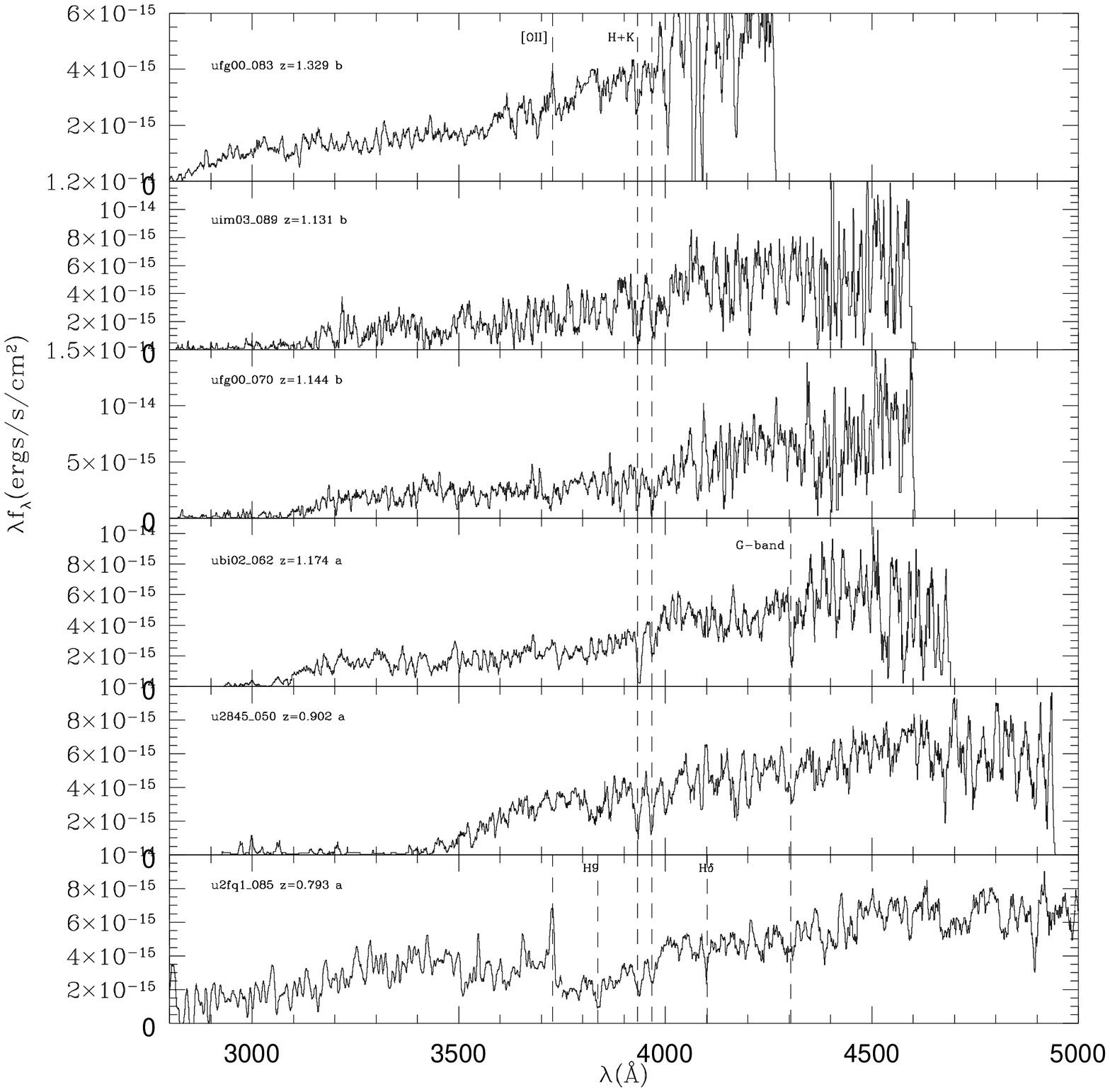}
\caption{A representative subsample of ERO spectra. We show
two spectra with emission line [OII] and four with pure absorption lines.
The spectra were smoothed with a boxcar of 11~pixels. 
\label{spec}}
\epsscale{1.0}
\end{figure}

\begin{figure}[!thp]
\plotone{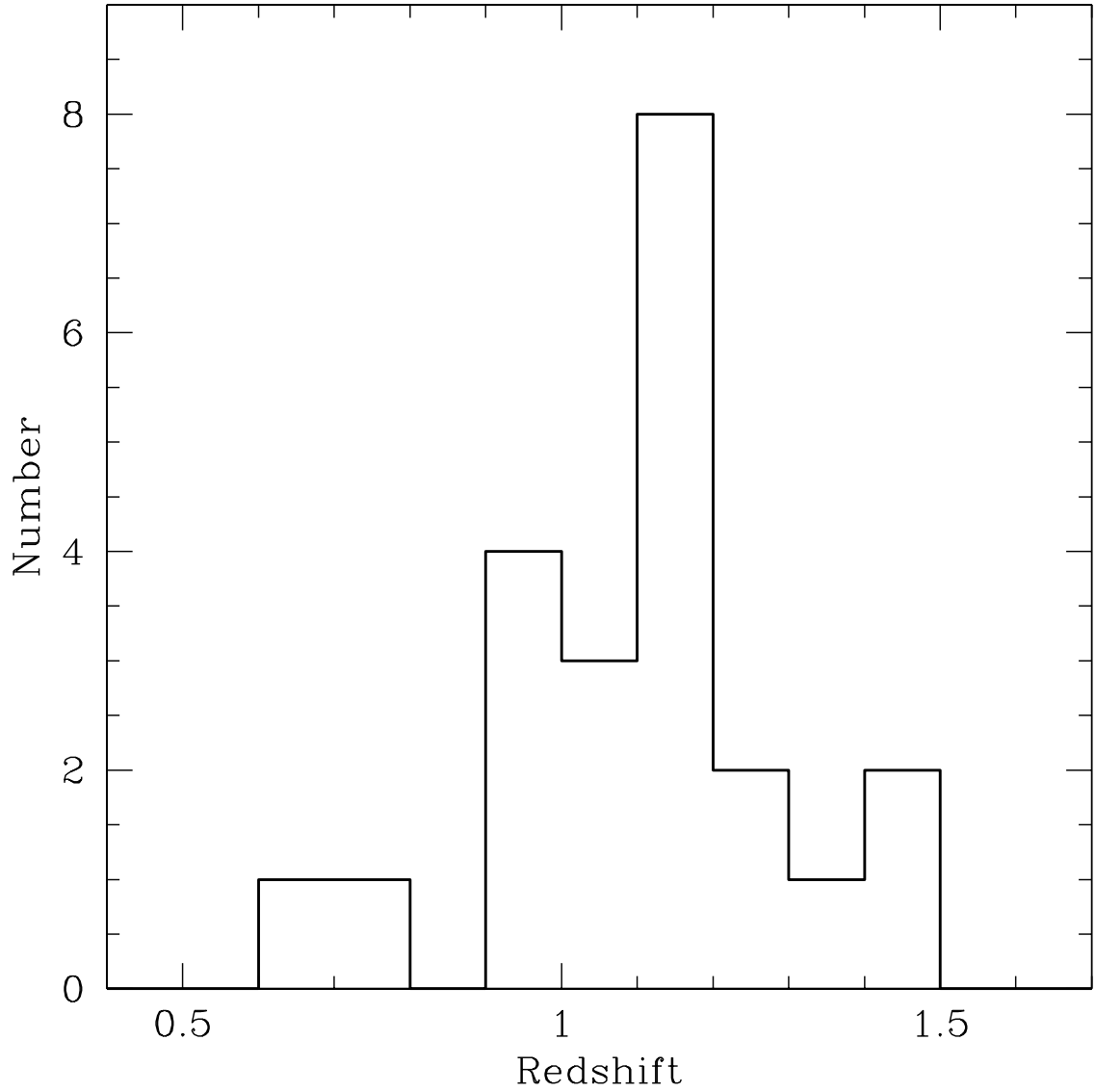}
\caption{The redshift distribution of our 24 sources. The median
redshift is 1.14. The bin size is 0.1. \label{zhist}}
\end{figure}

\begin{figure}[!thp]
\plotone{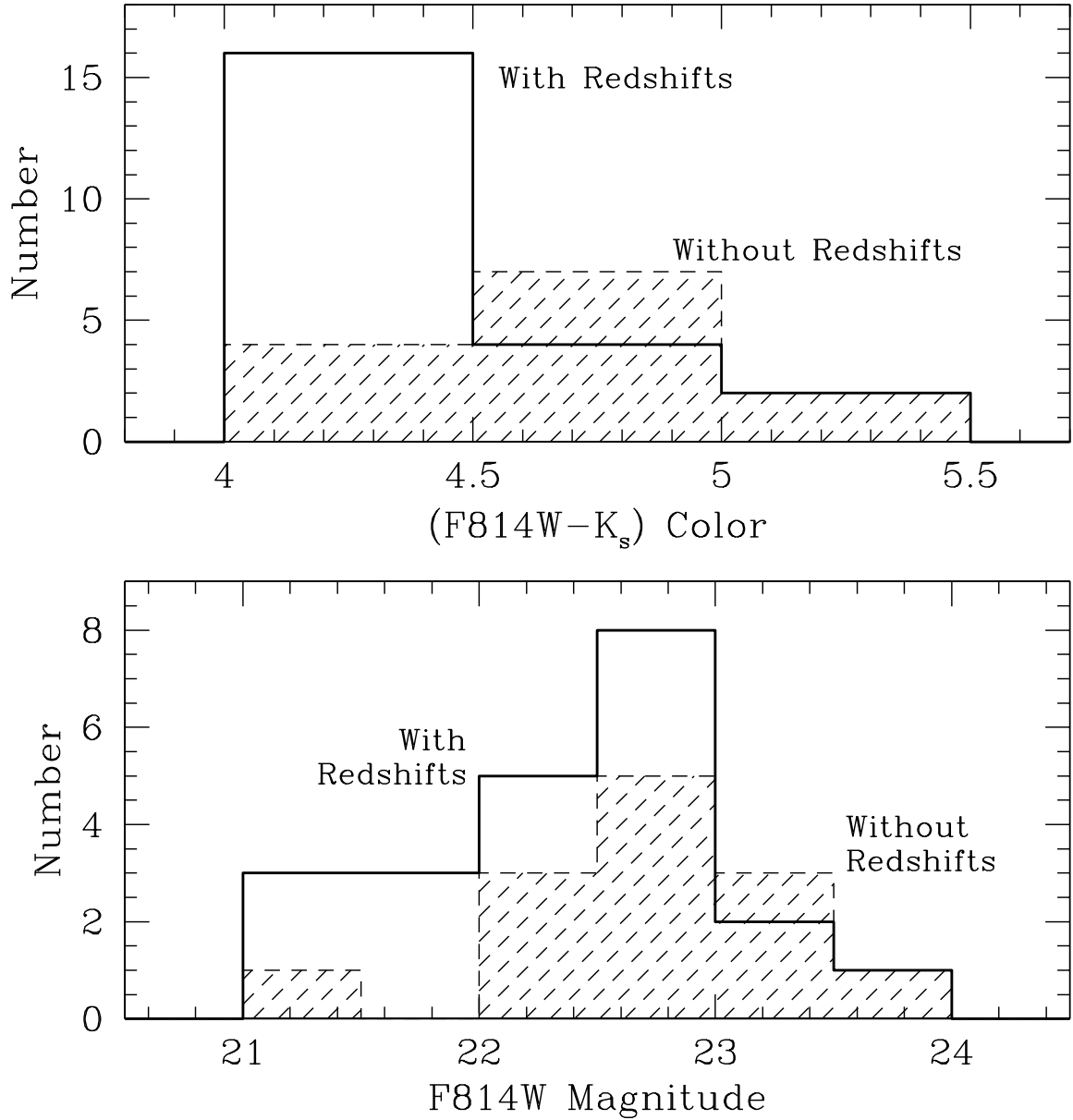}
\caption{The two panels show the distributions of the ($F814W - K_s$)
color (top) and $F814W$ magnitude (bottom) for the sources with (solid
lines) and without (dashed/shaded) redshifts.  \label{cmhist}}
\end{figure}

\begin{figure}[!thp]
\plotone{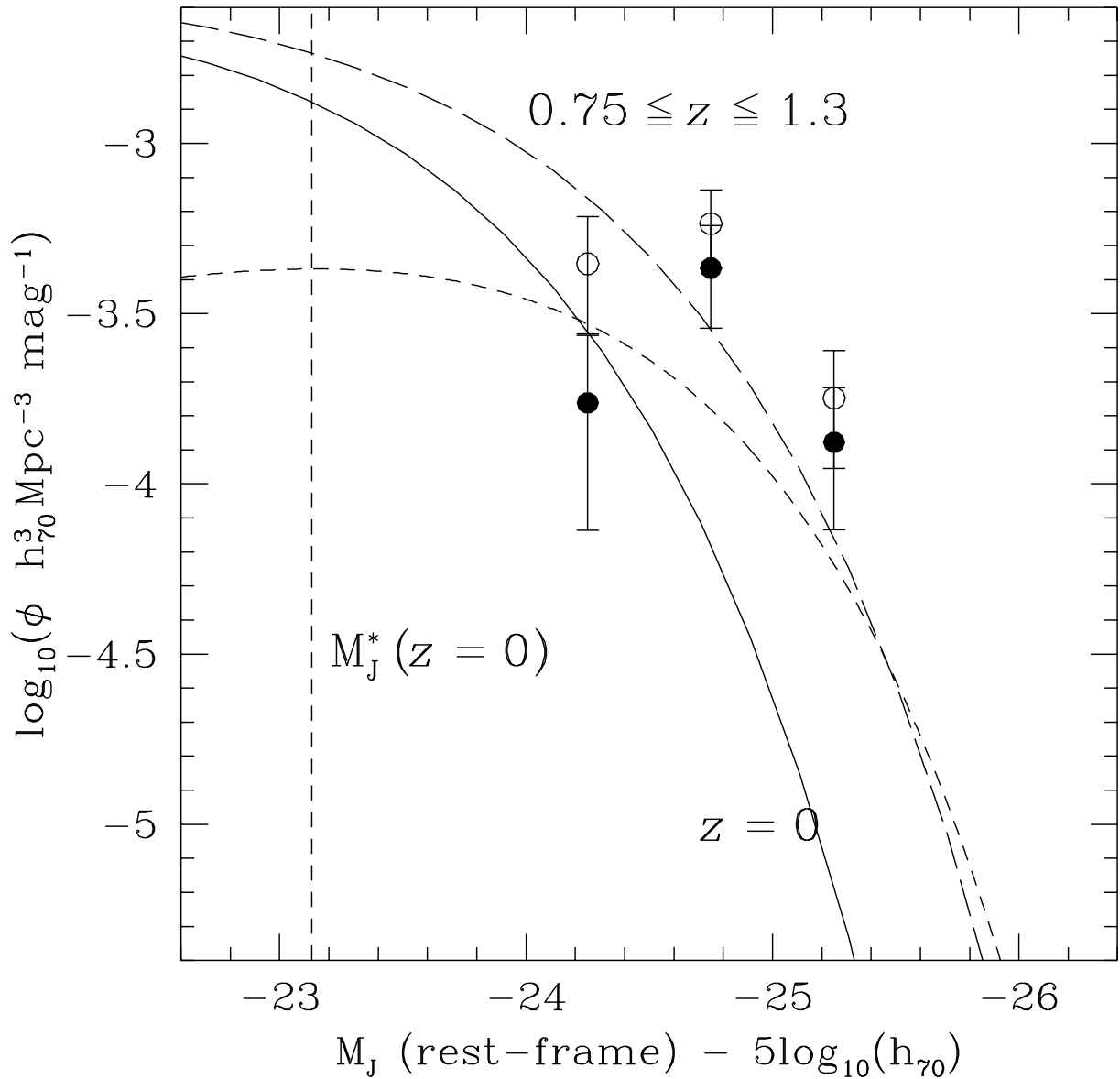}
\caption{The rest-frame J-band luminosity function. The three solid data 
points are from this survey using only the sources with spectroscopic redshifts, 
whereas the circles indicate the measurements including sources without
spectroscopic redshifts. The solid curve represents the local
J-band luminosity function from Cole et al. (2001). The dashed curve
is the rest-frame J-band luminosity function from Pozzetti et al. 
(2003) at the redshift range of $0.75 < z < 1.3$ with $z_{mean} = 1.0$, 
and the dotted line is also from the K20 survey (Pozzetti 2003, 
private communication), but for sources with pure absorption lines 
in the same redshift range.
\label{lf}}
\end{figure}

\begin{figure*}[!thp]
\epsscale{0.55}\plotone{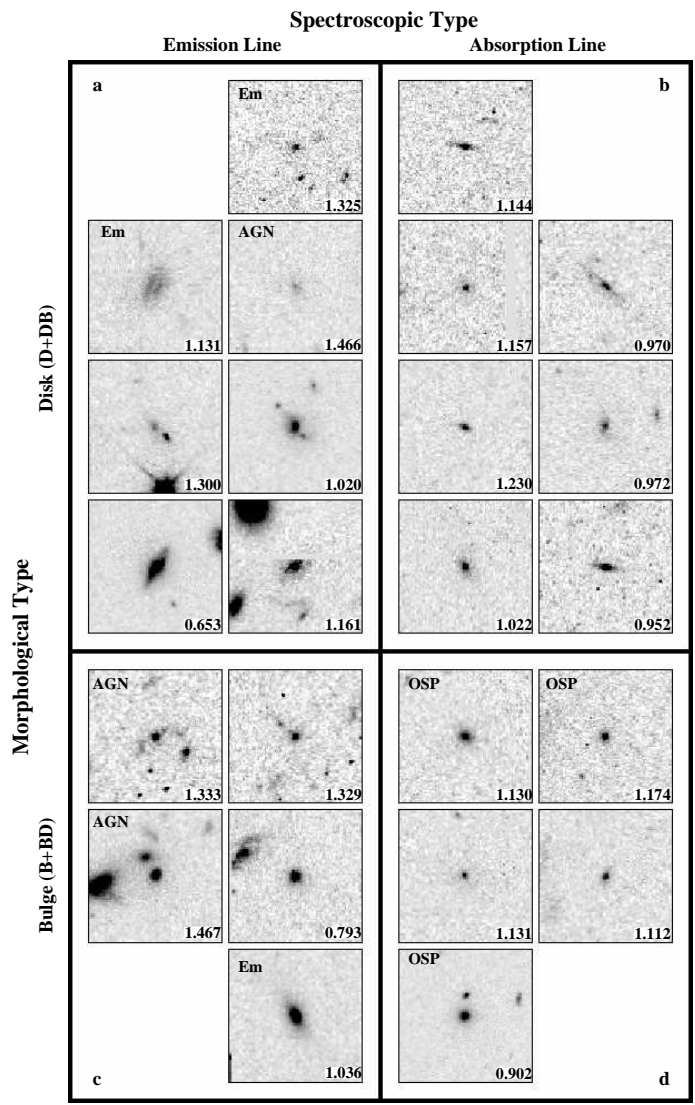}
\caption{The relationship between morphological and spectral types.  
The morphologies are divided between bulge or disk dominated, while 
the spectral types are divided by the presence or absence of emission 
lines.  The four groups of EROs are therefore (a) disks with emission 
lines, (b) disks without emission lines (only absorption lines), (c) 
bulges with emission lines, and (d) bulges without emission lines.  
The images are 8\farcs1 square, and marked with the source redshift.  
The three active galaxies (AGN), three pure old stellar populations 
(OSP), and three pure emission line disks (Em) are also marked, and 
discussed further in the text. 
\label{mosp}} 
\epsscale{1.0}
\end{figure*}

\begin{figure}[!thp]
\epsscale{0.8} \plotone{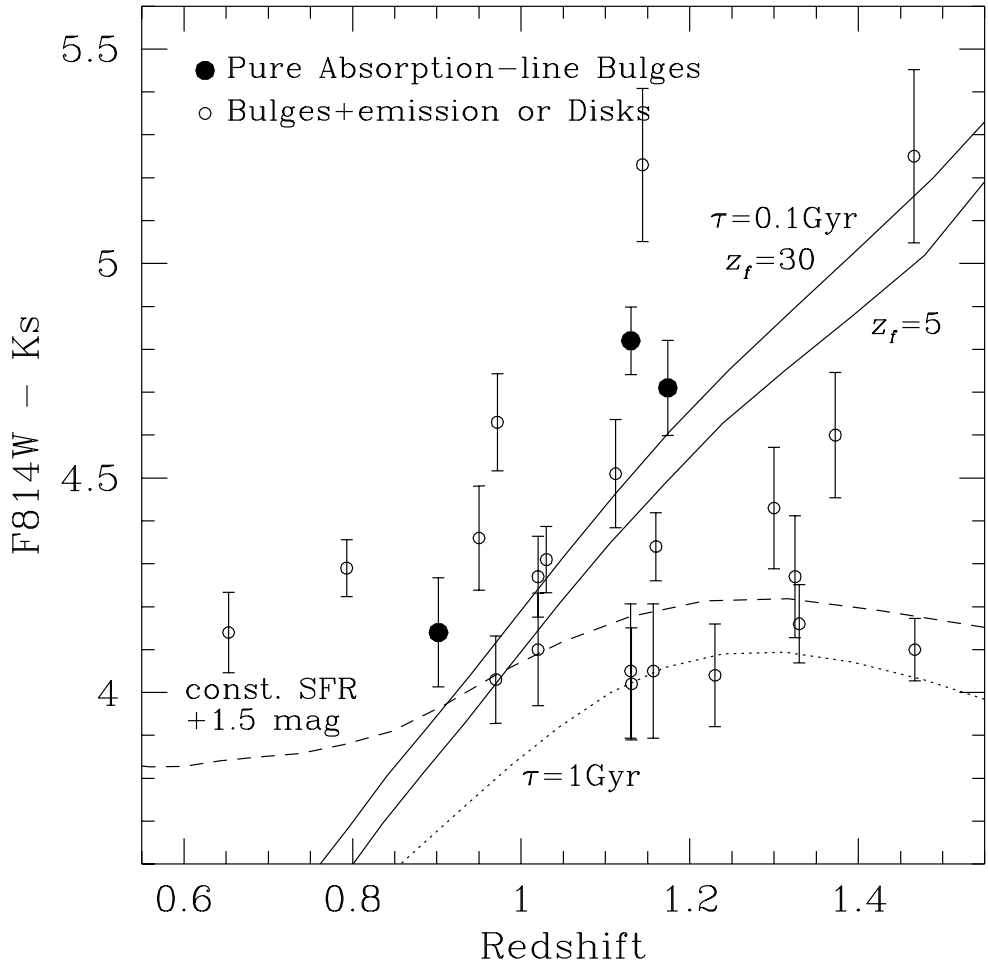}
\caption{The ($F814W - K_s$) color as a function of redshifts for the
sources listed in Table 1. The solid dots are the pure bulges with
absorption lines only, which could be isolated passively evolving
early-type galaxies. The open circles indicate the rest of EROs in the
sample, including bulges with small disks, bulges with emission lines
or disks. The curves are calculated from Bruzual \&\ Charlot (1996)
model SEDs, with the two solid curves for $\tau = 0.1$~Gyr (early-type
SED) with $z_{form} = 5$ and $z_{form}=30$ respectively. The dotted line
is for $\tau=1$~Gyr (late-type SED) and the dashed line for
constant star formation galaxies plus additional 1.5 magnitude of
extinction from dust in a simple foreground screen geometry.
\label{newcz}}
\epsscale{1.0}
\end{figure}

\begin{figure}[!thp]
\plotone{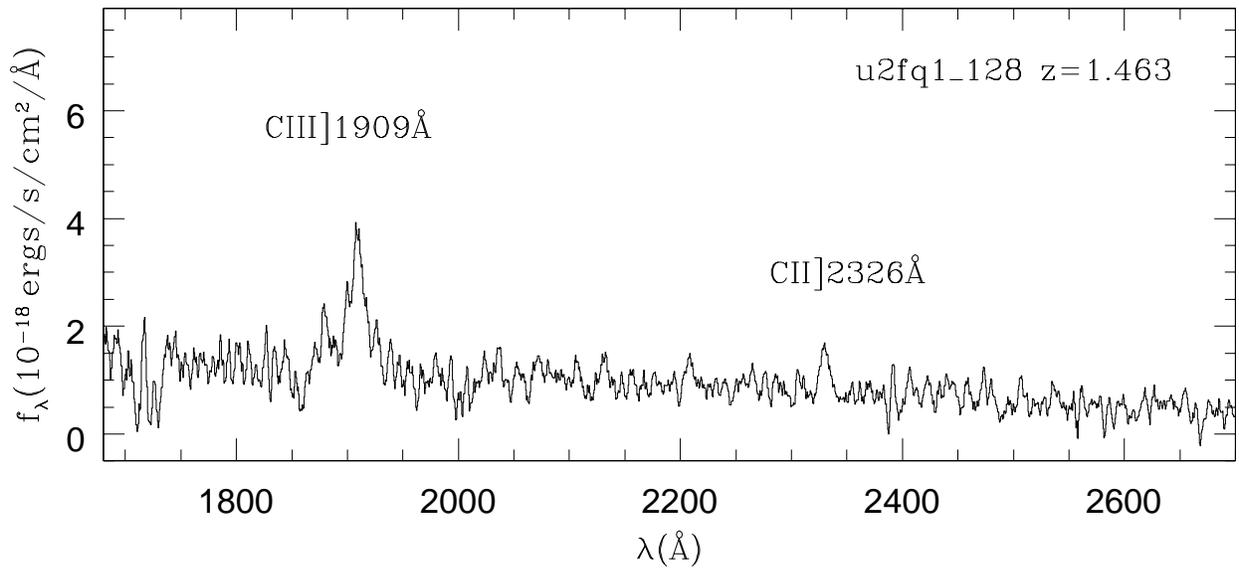}
\caption{This shows the AGN spectrum of u2fq1\_128.  The
\ion{C}{3}]~$\lambda$1909\AA\ line is 3740~km/s wide.  The spectrum is
smoothed with a boxcar of 5~pixels (7.3\AA).\label{agn}}
\end{figure}

\begin{figure}[!thp]
\epsscale{0.6} \plotone{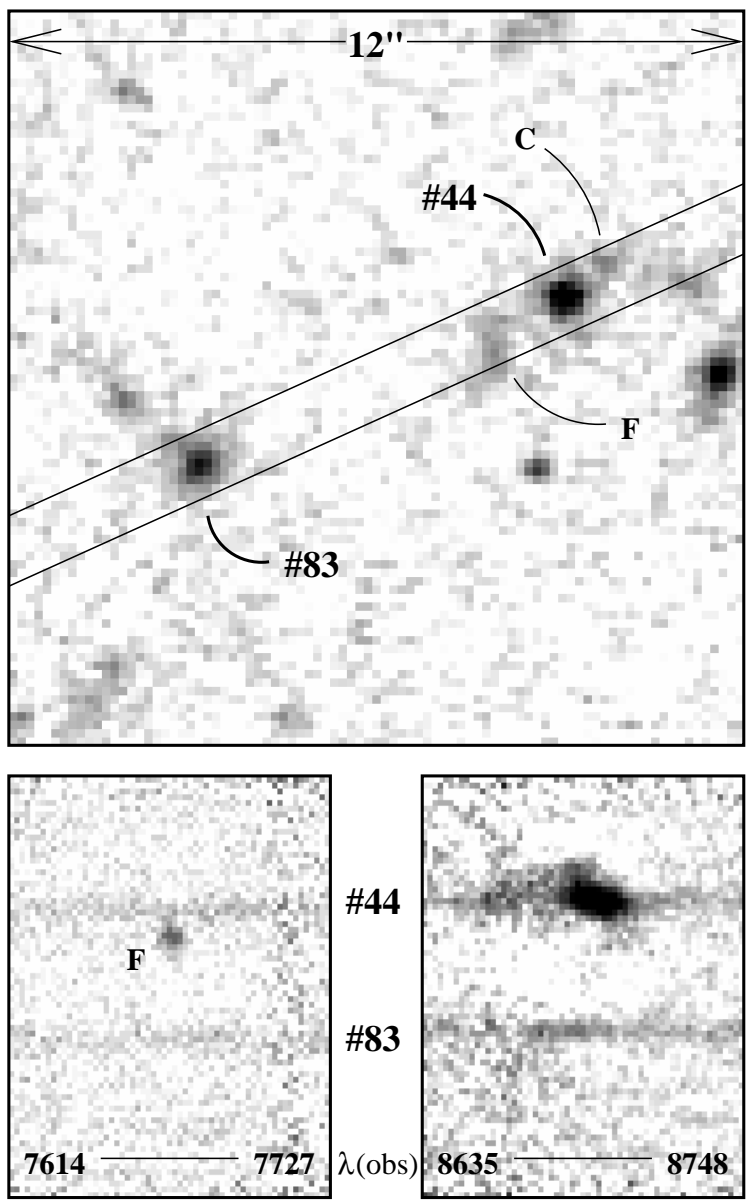}
\caption{The top panel shows an HST/WFPC2 F814W image of EROs
ufg00\_044 and ufg00\_083 over a region of 12$^{''}\times$12$^{''}$.
The 1\farcs2 wide slit (parallel lines) covered both EROs as well as
two fainter objects near ufg00\_044.  The bottom panels show 2D
spectra centered on the brightest emission lines, covering the
observed wavelength regions of 7614--7727\AA\ and 8635--8748\AA.  The
companion (C) galaxy is at the same redshift as the two EROs, while
the foreground (F) galaxy is at $z=0.50$.\label{agn2}} \epsscale{1.0}
\end{figure}

\clearpage
\begin{deluxetable}{rrcrcrll}
   \tablewidth{0pt}
   \tabletypesize{\scriptsize}
   \tablecaption{Table 1: Redshifts of the 24 EROs \label{tab_v2}}
   \tablehead{
      \colhead{ID}                 &
      \colhead{MDS ID}             &
      \colhead{$K_s$/Color}        &
      \colhead{z/Q\tnm{1}}         &
      \colhead{Flux\tnm{2}}        &
      \colhead{Morph\tnm{3}}       &
      \colhead{Emission\tnm{4}}    &
      \colhead{Absorption\tnm{4}}  }
   \startdata
     2 & ufg00\_070        & 18.27/5.23 & 1.144/b & \nd  & DB &                          & H+K                        \\
     3 & ufg00\_121        & 18.29/4.27 & 1.325/a & 3.90 & D  & [O II]                   & H+K                        \\
     4 & ufg00\_044        & 17.68/4.16 & 1.333/a & 15.1 & B  & [O II], [NeIII]          & H+K                        \\
     5 & ufg00\_083        & 18.05/4.60 & 1.329/b & 0.58 & B  & [O II]                   & H+K                        \\
     7 & udh00\_057        & 18.58/4.05 & 1.157/c & \nd  & D  &                          & 4000\AA?                   \\
     9 & ueh02\_063        & 17.43/4.82 & 1.130/b & \nd  & B  &                          & H+K, FeI+MgI               \\
    12 & ubi02\_055        & 18.49/4.03 & 0.970/c & \nd  & D  &                          & 4000\AA                    \\
    13 & ubi02\_062        & 18.46/4.71 & 1.174/a & \nd  & B  &                          & H+K, 4000\AA               \\
    32 & uim03\_102        & 18.60/4.04 & 1.230/b & \nd  & DB &                          & H+K, 4000\AA               \\
    33 & uim03\_089        & 18.57/4.02 & 1.131/b & \nd  & BD &                          & H+K, 4000\AA               \\
    34 & uim03\_100        & 18.34/4.51 & 1.112/b & \nd  & BD &                          & H+K, 4000\AA               \\
    35 & uim03\_075        & 17.94/4.63 & 0.972/a & \nd  & DB &                          & H+K, 4000\AA, G-band       \\
    78 & u2845\_050        & 17.95/4.14 & 0.902/a & \nd  & B  &                          & H+K, G-band, MgII          \\
    94 & u2fq1\_130        & 17.91/4.05 & 1.131/a & 1.40 & D  & [O II]                   &                            \\
    95 & u2fq1\_389        & 18.39/5.25 & 1.466/a & 1.85 & D  & [O II], CIII] CII]       & 2900\AA?                   \\
    97 & u2fq1\_273        & 18.15/4.43 & 1.300/a & 1.26 & ID & [O II]                   & H+K                        \\
    99 & u2fq1\_128        & 17.37/4.10 & 1.467/a & 1.72 & B  & [O II], MgII, CIII] CII] &                            \\
   104 & u2fq1\_085        & 17.04/4.29 & 0.793/a & 1.66 & BD & [O II]                   & H+K, H$\delta$, H9, G-band \\
   105 & u2fq1\_115        & 17.86/4.31 & 1.036/a & 0.56 & B  & [O II]                   &                            \\
   109 & u2h91\_034\tnm{5} & 17.70/4.27 & 1.020/a & \nd  & DB & [O II]                   &                            \\
   110 & u2h91\_011\tnm{5} & 17.25/4.14 & 0.653/a & \nd  & D  & [O II]                   &                            \\
   111 & u2fq2             & 17.58/4.34 & 1.161/a & 2.10 & D  & [O II]                   & H+K, MgII                  \\
   113 & uec00\_053        & 18.01/4.10 & 1.022/a & \nd  & DB &                          & H+K, MgII                  \\
   115 & uec00\_047        & 17.81/4.36 & 0.952/b & \nd  & D  &                          & H+K                        \\
   \enddata    
   \tablenotetext{1}{The quality of the redshift measurements are ranked from high to low
                     as a, b, and c.}
   \tablenotetext{2}{The [\ion{O}{2}]~$\lambda$3727\AA\ emission line fluxes are in units of 
                     $10^{-17}$~ergs/s/cm$^2$. The two sources whose redshifts were taken 
                     from the Hawaii Redshift Survey don't have published 
                     [\ion{O}{2}]$\lambda$3727\AA\ line fluxes.}
   \tablenotetext{3}{Morphology from Paper I: D$=$Disk, DB$=$Disk$+$Bulge, BD$=$Bulge$+$Disk, B$=$Bulge}
   \tablenotetext{4}{Emission features:   [O II]\,3727\AA, [NeIII]\,3867\AA, MgII\,2802.7\AA, C III], C II]; 
                     Absorption features: H$+$K$=$Ca II H and K lines, 4000\AA is the 4000\AA\ break, 
                     2900\AA is the 2900\AA\ break, FeI+MgI are absorption lines at 3840\AA\ and 3580\AA\, 
                     G-band\,4303\AA. }
   \tablenotetext{5}{These two redshifts are from the Hawaii redshift survey (Cowie et al. 1996).}
\end{deluxetable}
\clearpage

\end{document}